%% file: audit_logs_arxiv_csCY.tex
\pdfoutput=1
\documentclass[11pt]{article}

\usepackage[utf8]{inputenc}
\usepackage[T1]{fontenc}
\usepackage{lmodern}
\usepackage[margin=1in]{geometry}
\usepackage{microtype}
\usepackage[sort&compress,numbers]{natbib}
\usepackage{graphicx}
\usepackage{booktabs}
\usepackage{tabularx}
\usepackage{caption}
\usepackage{tikz}
\usetikzlibrary{arrows.meta,shapes.geometric,decorations.pathreplacing}
\usepackage[hidelinks]{hyperref}
\usepackage{authblk}

\newcommand{\missingfilebox}[1]{%
  \fbox{%
    \begin{minipage}[c][0.22\textheight][c]{0.94\linewidth}
      \centering\small #1
    \end{minipage}%
  }%
}

\newcommand{\auditmatrixfallback}{%
\begin{tabular}{@{}p{0.46\linewidth}cccc@{}}
\toprule
\textbf{Logged event} & \textbf{Clinician} & \textbf{Patient} & \textbf{Team} & \textbf{Process} \\
\midrule
Stroke alert initiated for Patient~A &  & $\bullet$ & $\bullet$ & $\bullet$ \\
Emergency physician opens Patient~B chart & $\bullet$ &  &  &  \\
Radiology result posted for Patient~A &  & $\bullet$ &  & $\bullet$ \\
Emergency physician opens CT angiography result for Patient~A & $\bullet$ & $\bullet$ & $\bullet$ & $\bullet$ \\
Nurse documents neurologic assessment for Patient~A &  & $\bullet$ & $\bullet$ & $\bullet$ \\
Treatment decision documented for Patient~A &  & $\bullet$ & $\bullet$ & $\bullet$ \\
\bottomrule
\end{tabular}%
}

\title{Clinical Audit Logs as Multi-Axial Traces of Care Delivery}

\author[1]{Braden Eberhard\thanks{Correspondence: \texttt{braden.eberhard@pennmedicine.upenn.edu}}}
\author[2]{Nate C. Apathy}
\author[1,3]{Kevin B. Johnson}

\affil[1]{Department of Biostatistics, Epidemiology, and Informatics, Perelman School of Medicine, University of Pennsylvania, Philadelphia, PA, USA}
\affil[2]{Department of Health Policy and Management, University of Maryland School of Public Health, College Park, MD, USA}
\affil[3]{Department of Computer and Information Science, School of Engineering and Applied Science, University of Pennsylvania, Philadelphia, PA, USA}

\date{}

\hypersetup{
  pdftitle={Clinical Audit Logs as Multi-Axial Traces of Care Delivery},
  pdfauthor={Braden Eberhard, Nate C. Apathy, Kevin B. Johnson},
}

\begin{document}

\maketitle

\begin{abstract}
Electronic health record audit logs record timestamped actions through which clinical work is carried out. Generated as operational metadata, they now support research on clinician effort, patient outcomes, care-team coordination, and workflow structure. This Perspective explains that breadth by articulating audit logs as multi-axial event streams and drawing implications for representation learning, evaluation, and governance. Each logged action belongs simultaneously to multiple clinically meaningful relations: a clinician's work, a patient's trajectory, a team's activity, and a recurring workflow. This structure motivates foundation-model pretraining to learn reusable representations over the raw stream. Reading audit logs as multi-axial traces specifies what such representations must preserve, how their value should be tested, and how their use should be governed.
\end{abstract}

\section{Introduction}
EHR audit logs are multi-axial: a single logged action is at once an act by a clinician, a step in a patient's trajectory, a contribution to a care team's work, and a step in a process. They are the behavioral trace of clinical work, the high-resolution, time-stamped record of the acts through which clinicians and staff access, navigate, modify, and document care in the EHR. In clinical care this structure is especially consequential, because these relations correspond to the units through which care is decided, delivered, and received.

Audit logs are generated automatically as a byproduct of clinical work. Originally retained for security and compliance, they have now become the basis for a rapidly growing body of health services research.\citep{AdlerMilstein2020Goldmine, Rule2020AuditLogReview, Kannampallil2023EarlyEfforts} The same raw stream has supported next-day discharge and deterioration prediction,\citep{Zhang2021NextDayDischarge, Rossetti2025CONCERN} the measurement of clinician effort and documentation burden,\citep{Kim2024ActionEntropy, Apathy2023NoteLength} the reconstruction of care-team coordination,\citep{Chen2017CollaborativeTeams, Yakusheva2025MetadataMiningTeams} and the discovery of workflow structure.\citep{Hribar2018WorkflowOptimization, Zhang2022InferringWorkflows} These literatures have largely developed independently, each appropriately scoped to its own question and reconstructing from the stream only what that question required.

That the same raw stream can be organized productively for so many different endpoints is an opportunity, since versatility across endpoints invites learning the stream's general structure before any one endpoint fixes its shape. This is the premise of foundation-model pretraining, in which a single representation is learned from raw, abundant data before any task is chosen, and then reused across tasks.\citep{Bommasani2021FoundationModels, Steinberg2021CLMBR} Audit logs are well-suited to this approach, and a representation pretrained over the stream could provide a reusable starting point for sparse, noisy, and heterogeneous audit-log tasks. Its advantage over task-specific models should be greatest for outcomes whose signal is distributed across multiple orderings of the stream rather than contained within one.

\section{Audit logs as multi-axial event streams}
Recent work helps make sense of this growing range. Yan et al.\citep{Yan2025DualLens} emphasize the value of EHR use metadata for clinical AI and point toward foundation models built on it, and Tawfik et al.\citep{Tawfik2025EmergingDomains} organize the measurement domains audit logs have begun to support. We build from these accounts by locating a common structure in the logged events themselves that organizes these domains around the event rather than the endpoint.

We use \emph{axis} to mean a way of ordering events in the stream from a single vantage point, such as one clinician's session or one patient's trajectory. Any single ordering is partial, foregrounding the events that bear on its vantage point and leaving the rest as background. We call audit logs multi-axial because a single event can be ordered in several ways at once. We suggest clinician, patient, team, and process axes because they correspond most closely to how care is decided, delivered, and received.

\begin{figure}[t]
\centering

\noindent\makebox[\linewidth][l]{\textbf{\large (a)}}

\IfFileExists{MetroMap.png}{%
  \includegraphics[width=\linewidth,height=0.42\textheight,keepaspectratio]{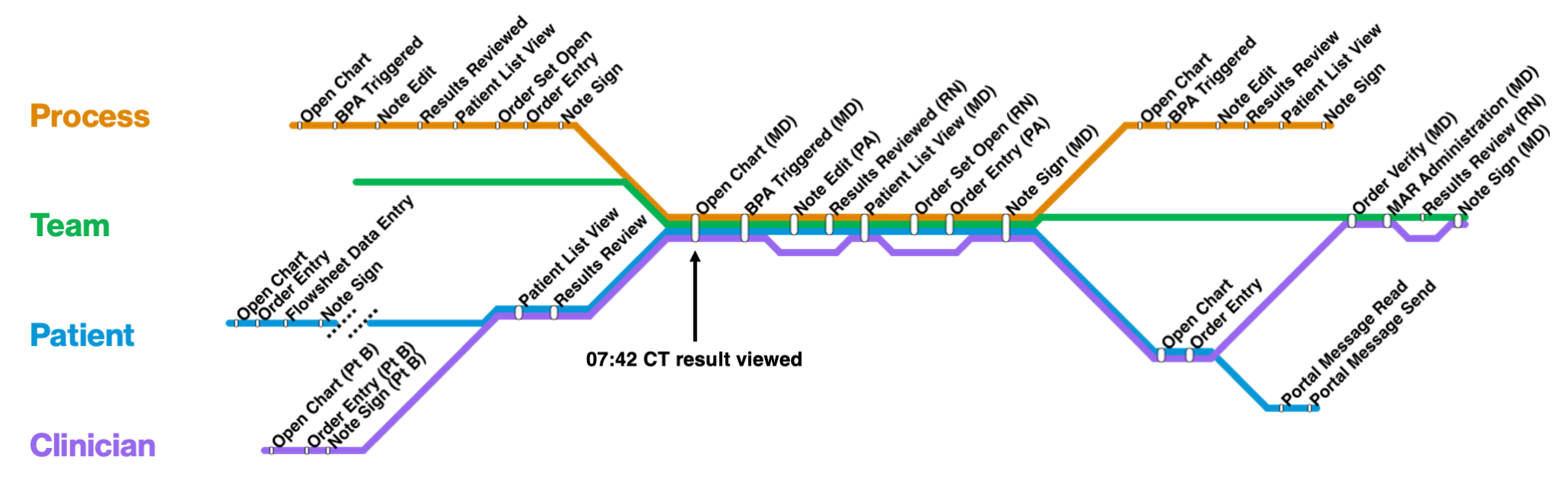}%
}{%
  \missingfilebox{MetroMap.png was not found. Upload the final figure file with this source, or replace this box with the intended panel.}%
}

\noindent\makebox[\linewidth][l]{\textbf{\large (b)}}

\IfFileExists{matrix.tex}{%
  \resizebox{\linewidth}{!}{\input{matrix.tex}}%
}{%
  \IfFileExists{matrix}{%
    \resizebox{\linewidth}{!}{\input{matrix}}%
  }{%
    \resizebox{\linewidth}{!}{\auditmatrixfallback}%
  }%
}

\caption{%
\textbf{Audit logs as a multi-axial substrate.}
We propose that a single logged action belongs to several clinically meaningful relations at once.
\textbf{(a)} A stroke-alert episode in which one anchor event is read along four axes: a step in the physician's session, a moment in the patient's evaluation after the alert and before treatment, one access within a window of overlapping chart use, and the imaging-review step of a time-sensitive protocol.
\textbf{(b)} A subset of the same raw events arranged in an event-by-axis matrix, with each axis foregrounding the relations that a different line of research has drawn on. A marker means the event participates in that axis; events with a single marker are background on the others. Patient~A is the stroke patient anchoring the episode, while Patients~B and~C are other patients the physician's session spans (MD, physician; RN, registered nurse; PA, physician assistant). The highlighted row is the anchor event of panel (a), the physician's 07:42 opening of the CT angiography result for Patient~A, traced through all four axes at once. A single representation pretrained over the full stream could serve these lines of work together, and could reach cross-axis questions.
}
\label{fig:multi_axial_event}
\end{figure}

Consider one audit log entry, traced through Figure~\ref{fig:multi_axial_event}, which renders the same stroke-alert episode two ways: panel (a) as a stylized transit map in which four lines (a clinician session, a patient trajectory, a care team, and a clinical process) run alongside one another and meet at shared events, and panel (b) as an event-by-axis matrix in which each logged action is scored for the axes it participates in. At 07:42, an emergency physician opens the CT angiography result for Patient A, six minutes after a stroke alert is initiated. In the raw audit log, this appears as a single row that may include a timestamp, the acting user and role, the patient in context, the action performed, the EHR object accessed, and the workstation or session. Its clinical meaning is inferred by reading that row in relation to the surrounding entries. A single audit log entry can therefore be situated simultaneously along four axes: the clinician work session, the patient trajectory, the care process, and the team workflow. Along the clinician axis, it marks an attention shift, a diagnostic step, or a response to new information. Along the patient axis, it falls within the stroke evaluation, after the alert and before a treatment decision. Along the process axis, it belongs to the imaging-review stage of a time-sensitive protocol. Along the team axis, it is one access among several overlapping interactions with the same chart, where the sequence and timing of who enters, reviews, orders, or documents becomes the unit of interest.

All four readings are carried by the same row, the shared substrate from which each line of research draws. A logged event joins a user, a patient, an object, a time, a role, and a workflow context, and each of these meanings is present in the record before any task selects among them. The axes run alongside one another and meet at shared events rather than in independent parallel tracks. Whether the 07:42 access reflects ordinary review or a coordination gap, for instance, depends on the protocol stage, the patient's acuity, and which team members had already reached the chart, an alignment visible only where the axes meet. A measure of task switching depends on the workflow context that makes a switch consequential,\citep{Bartek2023TaskSwitching} and a team reconstructed from co-access is shaped by the trajectories that bring users into the same chart.\citep{Rose2023TeamIsBrain} These four are not the only orderings a row supports, and workstation, chart object, or session may be central for particular questions.

\section{Multi-axiality as a pretraining hypothesis}
Audit logs meet the conditions under which foundation-model pretraining is worth considering. They are generated continuously and at scale, far exceeding the number of available labels. Individual entries acquire meaning from the surrounding context rather than from action labels alone. And many outcomes of interest are observed only sparsely, through surveys, chart review, or delayed endpoints. Together, these conditions invite a strategy that learns the structure of EHR-mediated clinical work from the full stream before any one endpoint is fixed.

Recent studies have begun demonstrating signal using pretrained models and embeddings learned from audit logs in cognitive effort, next-day discharge, physician burnout, macrostructure in EHR work, and next-action prediction over hundreds of millions of inpatient actions.\citep{Kim2024ActionEntropy, Zhang2025DischargeAuditLogs, Liu2022HiPAL, Lou2022BurnoutAuditLogs, Lou2023Macrostructure, Kim2026Language} A multi-axial reading adds a more specific prediction about where that value should concentrate. If the advantage of pretraining comes from learning patterns of context, timing, and co-occurrence across the full stream, then it should be most pronounced for targets that depend on relations among orderings, and more modest for those largely contained within one. Table~\ref{tab:audit_fm_impact} outlines several of these, spanning within-axis measures and questions that turn on the alignment among axes.

This sits within a broader shift toward pretraining over structured event streams. Patient-centered EHR foundation models such as CLMBR and MOTOR show that longitudinal records support reusable representations learned at scale,\citep{Steinberg2021CLMBR, Steinberg2024MOTOR} while behavior foundation models such as PinFM and BehaveGPT extend the premise to large-scale activity traces.\citep{Chen2025PinFM, Gong2025BehaveGPT} Audit logs share what makes both work, abundant unlabeled sequences whose structure is learnable independent of any downstream task, while recording a behavioral signal largely absent from patient-centered streams and dimensions of clinical work those models are not designed to represent.

\begin{table}[t]
\centering
\footnotesize
\renewcommand{\arraystretch}{1.3}
\setlength{\tabcolsep}{6pt}
\begin{tabularx}{\linewidth}{@{} p{0.24\linewidth} c X @{}}
\multicolumn{3}{@{}l}{\Large\textbf{Clinical use cases for a behavioral foundation model of the EHR}} \\
\addlinespace[0.3em]
\midrule
\textbf{Capability} & \textbf{Axes} & \textbf{What a shared representation reaches} \\
\midrule
Patient deterioration
 & Pt, C
 & Activity around a patient carries behavioral information about clinical state, supplementing the documented record in deterioration prediction and amplifying signal where outcome labels are too sparse to engineer features locally. \\
Documentation burden
 & C, P
 & Clinician effort interpreted within its context rather than counted in isolation, shaped by the stage of the workflow and the interruptions that determine when a task switch carries a cost. \\
Handoff integrity
 & P, Pt, T
 & Whether responsibility transfers cleanly between clinicians, recoverable from how one clinician's attention to a chart ends relative to when another's begins and whether pending results are picked up across that boundary. \\
Care team coordination
 & T
 & Coordination as a temporal pattern of who reaches a chart and when, across people and patients, grounded in the trajectories that bring users to the same record. \\
Equity in conduct of care
 & Pt, C, T
 & Differences in timing, attention, and escalation among clinically similar patients, observable in how care is delivered before they surface in outcomes. \\
\bottomrule
\end{tabularx}
\caption{\textbf{Capabilities a shared audit-log representation could support, with the axes each draws on.}
Each row names a capability and describes what a representation pretrained over the full stream recovers that task-specific pipelines built around a single axis are not designed to capture. Axes denote the orderings of the stream each capability relies on: Pt, patient; C, clinician; T, team; P, process. The rows span questions that rest largely on one axis as well as questions that turn on how several axes align.}
\label{tab:audit_fm_impact}
\end{table}

\section{Implementation considerations}

A representation that delivers on this claim should reflect the structure of the stream itself. Each event should be treated as more than an action label, and time should remain a carrier of clinical meaning through relative intervals, event order, and the boundaries of encounters and sessions. The representation also ought to travel across sites, where institutions differ in configurations, workflows, and the actions their systems record. A model that learns only local event names or screen paths would mistake a site's surface for a general pattern of care. The portable signal is more relational, lying in when actions occur, which roles take them, what precedes and follows them, and how activity gathers around patients, clinicians, teams, and processes. Done well, this would allow a site with little annotation to inherit a working signal rather than train only on its own labels, as shared EHR foundation models have been shown to transfer across institutions.\citep{Guo2024SharedFM}

The framing also implies what a benchmark for these representations would have to contain. Audit-log studies are at present evaluated in isolation, each on its own cohort, site, and outcome, so a representation or measure built at one institution is not easily compared against one built at another, and neither can be tested for the cross-axis structure the claim turns on. A benchmark able to test that claim would assemble a common set of tasks spanning axes and graded by their dependence on cross-axis structure, from within-axis tasks such as a clinician's active time, through mixed tasks such as next-day discharge, to cross-axis tasks such as coordination breakdown and inequity in the conduct of care. It would judge joint representations' performance against single-axis models, ensembles, and strong hand-engineered baselines on identical inputs drawn from the raw stream, and span multiple sites.

\section{Governance and equity}

The governance and equity stakes weigh more heavily on audit logs than on most clinical data, because audit logs record how care is delivered rather than only what was documented or what resulted. This gives them a complementary role in the study of inequity. Outcome and chart-review measures capture results; the stream also records the timing, attention, escalation, and coordination through which those results are produced, where disparity may appear before it reaches an outcome. A model sensitive enough to identify disparities in care delivery is also sensitive enough to encode and reproduce the institutional patterns, resource constraints, and biased behaviors embedded in that delivery. Because the property cuts both ways, governance must address how a model is used rather than the model alone. System-learning uses, which study workflows, teams, and processes, differ in kind from individual-evaluation uses aimed at performance assessment, discipline, or productivity scoring, which should be treated as high-risk and require separate justification and oversight.

\section{Conclusion}

Audit logs are usually treated as operational records or as a source of task-specific measures. We propose that they be read as multi-axial event streams, behavioral traces whose events carry several clinically meaningful relations at once, and that this reading motivates foundation-model pretraining over the stream. One representation could support tasks the field now approaches one at a time, spanning clinician effort, patient outcomes, team coordination, and workflow. Its advantage should be greatest on the questions that remain hard precisely because they cross axes or lack labels, and that prediction is what a shared, multi-site benchmark would test. Realizing this would require representations that preserve the contextual and temporal structure of events, a benchmark that grades tasks by their dependence on cross-axis structure, and governance that keeps such models pointed at systems of care rather than at individual surveillance. Read as a multi-axial substrate, audit logs offer a way to study care delivery through the trace it already leaves behind.

\bibliographystyle{unsrtnat}
\bibliography{references}

\end{document}

%% file: matrix.tex
\definecolor{Pt}{HTML}{0896D7}
\definecolor{process}{HTML}{DF8600}
\definecolor{team}{HTML}{00B251}
\definecolor{clinician}{HTML}{9768EE}
\definecolor{inkmuted}{HTML}{5F5E5A}
\definecolor{hifill}{HTML}{FAEEDA}
\definecolor{gridline}{HTML}{E6E2DA}


\begin{tikzpicture}[
    collbl/.style={text=inkmuted, align=center},
    rowlbl/.style={
      font=\fontsize{7.5pt}{8.5pt}\selectfont,
      text=inkmuted, anchor=west, align=left, text ragged, text width=4.4cm
    },
    rowlblA/.style={
      font=\fontsize{7.5pt}{8.5pt}\selectfont\bfseries,
      text=black!90, anchor=west, align=left, text ragged, text width=4.4cm
    },
    grplbl/.style={
      font=\fontsize{7pt}{8pt}\selectfont\itshape,
      text=inkmuted, anchor=south, align=center, rotate=90
    },
    dot/.style={circle, fill=#1, minimum size=8pt, inner sep=0},
    sq/.style={rectangle, fill=#1, minimum size=8pt, inner sep=0},
    tri/.style={isosceles triangle, isosceles triangle apex angle=60, fill=#1,
                minimum height=9pt, inner sep=0, shape border rotate=90},
    dia/.style={diamond, fill=#1, minimum width=10pt, minimum height=10pt, inner sep=0},
  ]

  \def\colP{6.15}
  \def\colC{8.35}
  \def\colT{10.55}
  \def\colPr{12.75}

  \def\rA{8.00}
  \def\rB{7.40}
  \def\rC{6.80}
  \def\rD{6.20}
  \def\rE{5.60}
  \def\rF{5.00}
  \def\rG{4.40}
  \def\rH{3.80}
  \def\rI{3.20}
  \def\rJ{2.60}
  \def\rK{2.00}
  \def\rL{1.40}
  \def\rM{0.80}
  \def\rN{0.20}
  \def\rO{-0.40}
  \def\rP{-1.00}

  \def\lblx{1.30}

  \fill[hifill, rounded corners=4pt]
    (1.15,\rK-0.30) rectangle (13.3,\rK+0.30);

  \foreach \y in {\rA,\rB,\rC,\rD,\rE,\rF,\rG,\rH,\rI,\rJ,\rK,\rL,\rM,\rN,\rO,\rP} {
    \draw[gridline, line width=0.35pt] (5.75,\y) -- (13.3,\y);
  }

  \node[collbl] at (\colP,8.55)  {Patient};
  \node[collbl] at (\colC,8.55)  {Clinician};
  \node[collbl] at (\colT,8.55)  {Team};
  \node[collbl] at (\colPr,8.55) {Process};

  \node[rowlbl]  at (\lblx,\rA) {Open Chart};
  \node[rowlbl]  at (\lblx,\rB) {BPA Triggered};
  \node[rowlbl]  at (\lblx,\rC) {Order Entry};
  \node[rowlbl]  at (\lblx,\rD) {Open Chart (Pt B)};
  \node[rowlbl]  at (\lblx,\rE) {Care Plan Edit};
  \node[rowlbl]  at (\lblx,\rF) {Portal Message Send};
  \node[rowlbl]  at (\lblx,\rG) {Problem List Update};
  \node[rowlbl]  at (\lblx,\rH) {Order Verify (MD)};
  \node[rowlbl]  at (\lblx,\rI) {MAR Administration (MD)};
  \node[rowlbl]  at (\lblx,\rJ) {Results Review (RN)};
  \node[rowlblA] at (\lblx,\rK) {07:42 Open Chart (MD)};
  \node[rowlbl]  at (\lblx,\rL) {Patient List View (MD)};
  \node[rowlbl]  at (\lblx,\rM) {Flow Sheet Data Entry (MD)};
  \node[rowlbl]  at (\lblx,\rN) {Order Entry (RN)};
  \node[rowlbl]  at (\lblx,\rO) {Note Edit (PA)};
  \node[rowlbl]  at (\lblx,\rP) {Order Modify (Pt C)};

  \def\brx{0.55}
  \def\lblgx{0.28}
  \tikzset{grpbrace/.style={decorate, decoration={brace, amplitude=5pt, raise=1pt}, line width=0.7pt, inkmuted}}

  \draw[grpbrace] (\brx,\rC-0.28) -- (\brx,\rA+0.28);
  \node[grplbl] at (\lblgx,{(\rA+\rC)/2}) {Recurring process};

  \draw[grpbrace] (\brx,\rG-0.28) -- (\brx,\rE+0.28);
  \node[grplbl] at (\lblgx,{(\rE+\rG)/2}) {Prior visit};

  \draw[grpbrace] (\brx,\rJ-0.28) -- (\brx,\rH+0.28);
  \node[grplbl, text width=2.2cm] at (\lblgx,{(\rH+\rJ)/2}) {Other process,\\same team};

  \draw[grpbrace] (\brx,\rO-0.28) -- (\brx,\rK+0.28);
  \node[grplbl] at (\lblgx,{(\rK+\rO)/2}) {Active visit};

  \node[dia=process] at (\colPr,\rA){};
  \node[dia=process] at (\colPr,\rB){};
  \node[dia=process] at (\colPr,\rC){};

  \node[sq=clinician] at (\colC,\rD){};
  \node[dot=Pt] at (\colP,\rE){};

  \node[dot=Pt]       at (\colP,\rF){};
  \node[sq=clinician] at (\colC,\rF){};
  \node[dot=Pt]       at (\colP,\rG){};
  \node[sq=clinician] at (\colC,\rG){};

  \node[sq=clinician] at (\colC,\rH){};
  \node[tri=team]     at (\colT,\rH){};
  \node[sq=clinician] at (\colC,\rI){};
  \node[tri=team]     at (\colT,\rI){};
  \node[sq=clinician] at (\colC,\rJ){};
  \node[tri=team]     at (\colT,\rJ){};

  \node[dot=Pt]       at (\colP,\rK){};
  \node[sq=clinician] at (\colC,\rK){};
  \node[tri=team]     at (\colT,\rK){};
  \node[dia=process]  at (\colPr,\rK){};
  \node[dot=Pt]       at (\colP,\rL){};
  \node[sq=clinician] at (\colC,\rL){};
  \node[tri=team]     at (\colT,\rL){};
  \node[dia=process]  at (\colPr,\rL){};
  \node[dot=Pt]       at (\colP,\rM){};
  \node[sq=clinician] at (\colC,\rM){};
  \node[tri=team]     at (\colT,\rM){};
  \node[dia=process]  at (\colPr,\rM){};

  \node[dot=Pt]      at (\colP,\rN){};
  \node[tri=team]    at (\colT,\rN){};
  \node[dia=process] at (\colPr,\rN){};
  \node[dot=Pt]      at (\colP,\rO){};
  \node[tri=team]    at (\colT,\rO){};
  \node[dia=process] at (\colPr,\rO){};

  \node[sq=clinician] at (\colC,\rP){};

\end{tikzpicture}